%
%
%
%
%
%
%
%
\def\standardrisposta{s }\def\reducedrisposta{r }
\def\mplarisposta{mpla }\def\zerorisposta{z }
\def\doublerisposta{d }\def\cartarisposta{e }\def\amsrisposta{y }
\newcount\ingrandimento \newcount\sinnota \newcount\dimnota
\newcount\unoduecol \newdimen\collhsize \newdimen\tothsize
\newdimen\fullhsize \newcount\controllorisposta \sinnota=1
\newskip\infralinea  \global\controllorisposta=0
%
%
%
%
%
\def\risposta{s }
\def\srisposta{e }
\def\arisposta{y }
\ifx\risposta\standardrisposta \ingrandimento=1200
\message {>> This will come out UNREDUCED << }
\dimnota=2 \unoduecol=1 \global\controllorisposta=1 \fi
\ifx\risposta\reducedrisposta \ingrandimento=1095 \dimnota=1
\unoduecol=1  \global\controllorisposta=1
\message {>> This will come out REDUCED << } \fi
\ifx\risposta\doublerisposta \ingrandimento=1000 \dimnota=2
\unoduecol=2   \message {>> You must print this in
LANDSCAPE orientation << } \global\controllorisposta=1 \fi
\ifx\risposta\mplarisposta \ingrandimento=1000 \dimnota=1
\message {>> Mod. Phys. Lett. A format << }
\unoduecol=1 \global\controllorisposta=1 \fi
\ifx\risposta\zerorisposta \ingrandimento=1000 \dimnota=2
\message {>> Zero Magnification format << }
\unoduecol=1 \global\controllorisposta=1 \fi
\ifnum\controllorisposta=0  \ingrandimento=1200
\message {>>> ERROR IN INPUT, I ASSUME STANDARD
UNREDUCED FORMAT <<< }  \dimnota=2 \unoduecol=1 \fi
\magnification=\ingrandimento
%
%
%
%
\newdimen\eucolumnsize \newdimen\eudoublehsize \newdimen\eudoublevsize
\newdimen\uscolumnsize \newdimen\usdoublehsize \newdimen\usdoublevsize
\newdimen\eusinglehsize \newdimen\eusinglevsize \newdimen\ussinglehsize
\newskip\standardbaselineskip \newdimen\ussinglevsize
\newskip\reducedbaselineskip \newskip\doublebaselineskip
\eucolumnsize=12.0truecm    
\eudoublehsize=25.5truecm   
\eudoublevsize=6.5truein    
\uscolumnsize=4.4truein     
\usdoublehsize=9.4truein    
\usdoublevsize=6.8truein    
\eusinglehsize=6.5truein    
\eusinglevsize=24truecm     
\ussinglehsize=6.5truein    
\ussinglevsize=8.9truein    
\standardbaselineskip=16pt plus.2pt  
\reducedbaselineskip=14pt plus.2pt   
\doublebaselineskip=12pt plus.2pt    
%
%
\def\Portoffset{}
\def\Landoffset{}
\ifx\risposta\mplarisposta \def\Portoffset{\hoffset=1.8truecm} \fi
%
%
\def\Landspec{}
\tolerance=10000
\parskip=0pt plus2pt  \leftskip=0pt \rightskip=0pt
%
%
\ifx\risposta\standardrisposta \infralinea=\standardbaselineskip \fi
\ifx\risposta\reducedrisposta  \infralinea=\reducedbaselineskip \fi
\ifx\risposta\doublerisposta   \infralinea=\doublebaselineskip \fi
\ifx\risposta\mplarisposta     \infralinea=13pt \fi
\ifx\risposta\zerorisposta     \infralinea=12pt plus.2pt\fi
\ifnum\controllorisposta=0    \infralinea=\standardbaselineskip \fi
\ifx\risposta\doublerisposta   \Landoffset \else \Portoffset \fi
\ifx\risposta\doublerisposta \ifx\srisposta\cartarisposta
\tothsize=\eudoublehsize \collhsize=\eucolumnsize
\vsize=\eudoublevsize  \else  \tothsize=\usdoublehsize
\collhsize=\uscolumnsize \vsize=\usdoublevsize \fi \else
\ifx\srisposta\cartarisposta \tothsize=\eusinglehsize
\vsize=\eusinglevsize \else  \tothsize=\ussinglehsize
\vsize=\ussinglevsize \fi \collhsize=4.4truein \fi
\ifx\risposta\mplarisposta \tothsize=5.0truein
\vsize=7.8truein \collhsize=4.4truein \fi
%
%
%
%
\newcount\contaeuler \newcount\contacyrill \newcount\contaams
\font\ninerm=cmr9  \font\eightrm=cmr8  \font\sixrm=cmr6
\font\ninei=cmmi9  \font\eighti=cmmi8  \font\sixi=cmmi6
\font\ninesy=cmsy9  \font\eightsy=cmsy8  \font\sixsy=cmsy6
\font\ninebf=cmbx9  \font\eightbf=cmbx8  \font\sixbf=cmbx6
\font\ninett=cmtt9  \font\eighttt=cmtt8  \font\nineit=cmti9
\font\eightit=cmti8 \font\ninesl=cmsl9  \font\eightsl=cmsl8
\skewchar\ninei='177 \skewchar\eighti='177 \skewchar\sixi='177
\skewchar\ninesy='60 \skewchar\eightsy='60 \skewchar\sixsy='60
\hyphenchar\ninett=-1 \hyphenchar\eighttt=-1 \hyphenchar\tentt=-1
\def\bfmath{\cmmib}                 
\font\tencmmib=cmmib10  \newfam\cmmibfam  \skewchar\tencmmib='177
\font\tencmbsy=cmbsy10  \newfam\cmbsyfam  \skewchar\tencmbsy='60
\def\scaps{\cmcsc}                 
\font\tencmcsc=cmcsc10  \newfam\cmcscfam
\ifnum\ingrandimento=1095

\font\capsone=cmcsc10 at 10.95pt 

\else

\font\capsone=cmcsc10 at 12pt 
\fi

\def\ttaarr{\bf}		
\def\ppaarr{\sl}		

%
%
%
\newfam\eufmfam \newfam\msamfam \newfam\msbmfam \newfam\eufbfam
\def\Loadeulerfonts{\global\contaeuler=1 \ifx\arisposta\amsrisposta
\font\teneufm=eufm10              
\font\eighteufm=eufm8 \font\nineeufm=eufm9 \font\sixeufm=eufm6
\font\seveneufm=eufm7  \font\fiveeufm=eufm5
\font\teneufb=eufb10              
\font\eighteufb=eufb8 \font\nineeufb=eufb9 \font\sixeufb=eufb6
\font\seveneufb=eufb7  \font\fiveeufb=eufb5
\font\teneurm=eurm10              
\font\eighteurm=eurm8 \font\nineeurm=eurm9
\font\teneurb=eurb10              
\font\eighteurb=eurb8 \font\nineeurb=eurb9
\font\teneusm=eusm10              
\font\eighteusm=eusm8 \font\nineeusm=eusm9
\font\teneusb=eusb10              
\font\eighteusb=eusb8 \font\nineeusb=eusb9
\else \def\eufm{\tt} \def\eufb{\tt} \def\eurm{\tt} \def\eurb{\tt}
\def\eusm{\tt} \def\eusb{\tt}    \fi}
\def\loadeuler{\Loadeulerfonts\tenpoint}
\def\loadamsmath{\global\contaams=1 \ifx\arisposta\amsrisposta
\font\tenmsam=msam10 \font\ninemsam=msam9 \font\eightmsam=msam8
\font\sevenmsam=msam7 \font\sixmsam=msam6 \font\fivemsam=msam5
\font\tenmsbm=msbm10 \font\ninemsbm=msbm9 \font\eightmsbm=msbm8
\font\sevenmsbm=msbm7 \font\sixmsbm=msbm6 \font\fivemsbm=msbm5
\else \def\msbm{\bf} \fi \def\Bbb{\msbm} \def\symbl{\msam} \tenpoint}
\def\loadcyrill{\global\contacyrill=1 \ifx\arisposta\amsrisposta
\font\tenwncyr=wncyr10 \font\ninewncyr=wncyr9 \font\eightwncyr=wncyr8
\font\tenwncyb=wncyr10 \font\ninewncyb=wncyr9 \font\eightwncyb=wncyr8
\font\tenwncyi=wncyr10 \font\ninewncyi=wncyr9 \font\eightwncyi=wncyr8
\else \def\cyrill{\sl} \def\cyrilb{\sl} \def\cyrili{\sl} \fi\tenpoint}
\ifx\arisposta\amsrisposta
\font\sevenex=cmex7               
\font\eightex=cmex8  \font\nineex=cmex9
\font\ninecmmib=cmmib9   \font\eightcmmib=cmmib8
\font\sevencmmib=cmmib7 \font\sixcmmib=cmmib6
\font\fivecmmib=cmmib5   \skewchar\ninecmmib='177
\skewchar\eightcmmib='177  \skewchar\sevencmmib='177
\skewchar\sixcmmib='177   \skewchar\fivecmmib='177
\font\ninecmbsy=cmbsy9    \font\eightcmbsy=cmbsy8
\font\sevencmbsy=cmbsy7  \font\sixcmbsy=cmbsy6
\font\fivecmbsy=cmbsy5   \skewchar\ninecmbsy='60
\skewchar\eightcmbsy='60  \skewchar\sevencmbsy='60
\skewchar\sixcmbsy='60    \skewchar\fivecmbsy='60
\font\ninecmcsc=cmcsc9    \font\eightcmcsc=cmcsc8     \else
\def\cmmib{\fam\cmmibfam\tencmmib}\textfont\cmmibfam=\tencmmib
\scriptfont\cmmibfam=\tencmmib \scriptscriptfont\cmmibfam=\tencmmib
\def\cmbsy{\fam\cmbsyfam\tencmbsy} \textfont\cmbsyfam=\tencmbsy
\scriptfont\cmbsyfam=\tencmbsy \scriptscriptfont\cmbsyfam=\tencmbsy
\scriptfont\cmcscfam=\tencmcsc \scriptscriptfont\cmcscfam=\tencmcsc
\def\cmcsc{\fam\cmcscfam\tencmcsc} \textfont\cmcscfam=\tencmcsc \fi
\catcode`@=11
\newskip\ttglue
\gdef\tenpoint{\def\rm{\fam0\tenrm}
  \textfont0=\tenrm \scriptfont0=\sevenrm \scriptscriptfont0=\fiverm
  \textfont1=\teni \scriptfont1=\seveni \scriptscriptfont1=\fivei
  \textfont2=\tensy \scriptfont2=\sevensy \scriptscriptfont2=\fivesy
  \textfont3=\tenex \scriptfont3=\tenex \scriptscriptfont3=\tenex
  \def\mcal{\fam2 \tensy}  \def\mmit{\fam1 \teni}
  \textfont\itfam=\tenit \def\it{\fam\itfam\tenit}
  \textfont\slfam=\tensl \def\sl{\fam\slfam\tensl}
  \textfont\ttfam=\tentt \scriptfont\ttfam=\eighttt
  \scriptscriptfont\ttfam=\eighttt  \def\tt{\fam\ttfam\tentt}
  \textfont\bffam=\tenbf \scriptfont\bffam=\sevenbf
  \scriptscriptfont\bffam=\fivebf \def\bf{\fam\bffam\tenbf}
     \ifx\arisposta\amsrisposta    \ifnum\contaeuler=1
  \textfont\eufmfam=\teneufm \scriptfont\eufmfam=\seveneufm
  \scriptscriptfont\eufmfam=\fiveeufm \def\eufm{\fam\eufmfam\teneufm}
  \textfont\eufbfam=\teneufb \scriptfont\eufbfam=\seveneufb
  \scriptscriptfont\eufbfam=\fiveeufb \def\eufb{\fam\eufbfam\teneufb}
  \def\eurm{\teneurm} \def\eurb{\teneurb} \def\eusm{\teneusm}
  \def\eusb{\teneusb}    \fi    \ifnum\contaams=1
  \textfont\msamfam=\tenmsam \scriptfont\msamfam=\sevenmsam
  \scriptscriptfont\msamfam=\fivemsam \def\msam{\fam\msamfam\tenmsam}
  \textfont\msbmfam=\tenmsbm \scriptfont\msbmfam=\sevenmsbm
  \scriptscriptfont\msbmfam=\fivemsbm \def\msbm{\fam\msbmfam\tenmsbm}
     \fi      \ifnum\contacyrill=1     \def\cyrill{\tenwncyr}
  \def\cyrilb{\tenwncyb}  \def\cyrili{\tenwncyi}         \fi
  \textfont3=\tenex \scriptfont3=\sevenex \scriptscriptfont3=\sevenex
  \def\cmmib{\fam\cmmibfam\tencmmib} \scriptfont\cmmibfam=\sevencmmib
  \textfont\cmmibfam=\tencmmib  \scriptscriptfont\cmmibfam=\fivecmmib
  \def\cmbsy{\fam\cmbsyfam\tencmbsy} \scriptfont\cmbsyfam=\sevencmbsy
  \textfont\cmbsyfam=\tencmbsy  \scriptscriptfont\cmbsyfam=\fivecmbsy
  \def\cmcsc{\fam\cmcscfam\tencmcsc} \scriptfont\cmcscfam=\eightcmcsc
  \textfont\cmcscfam=\tencmcsc \scriptscriptfont\cmcscfam=\eightcmcsc
     \fi            \tt \ttglue=.5em plus.25em minus.15em
  \normalbaselineskip=12pt
  \setbox\strutbox=\hbox{\vrule height8.5pt depth3.5pt width0pt}
  \let\sc=\eightrm \let\big=\tenbig   \normalbaselines
  \baselineskip=\infralinea  \rm}
\gdef\ninepoint{\def\rm{\fam0\ninerm}
  \textfont0=\ninerm \scriptfont0=\sixrm \scriptscriptfont0=\fiverm
  \textfont1=\ninei \scriptfont1=\sixi \scriptscriptfont1=\fivei
  \textfont2=\ninesy \scriptfont2=\sixsy \scriptscriptfont2=\fivesy
  \textfont3=\tenex \scriptfont3=\tenex \scriptscriptfont3=\tenex
  \def\mcal{\fam2 \ninesy}  \def\mmit{\fam1 \ninei}
  \textfont\itfam=\nineit \def\it{\fam\itfam\nineit}
  \textfont\slfam=\ninesl \def\sl{\fam\slfam\ninesl}
  \textfont\ttfam=\ninett \scriptfont\ttfam=\eighttt
  \scriptscriptfont\ttfam=\eighttt \def\tt{\fam\ttfam\ninett}
  \textfont\bffam=\ninebf \scriptfont\bffam=\sixbf
  \scriptscriptfont\bffam=\fivebf \def\bf{\fam\bffam\ninebf}
     \ifx\arisposta\amsrisposta  \ifnum\contaeuler=1
  \textfont\eufmfam=\nineeufm \scriptfont\eufmfam=\sixeufm
  \scriptscriptfont\eufmfam=\fiveeufm \def\eufm{\fam\eufmfam\nineeufm}
  \textfont\eufbfam=\nineeufb \scriptfont\eufbfam=\sixeufb
  \scriptscriptfont\eufbfam=\fiveeufb \def\eufb{\fam\eufbfam\nineeufb}
  \def\eurm{\nineeurm} \def\eurb{\nineeurb} \def\eusm{\nineeusm}
  \def\eusb{\nineeusb}     \fi   \ifnum\contaams=1
  \textfont\msamfam=\ninemsam \scriptfont\msamfam=\sixmsam
  \scriptscriptfont\msamfam=\fivemsam \def\msam{\fam\msamfam\ninemsam}
  \textfont\msbmfam=\ninemsbm \scriptfont\msbmfam=\sixmsbm
  \scriptscriptfont\msbmfam=\fivemsbm \def\msbm{\fam\msbmfam\ninemsbm}
     \fi       \ifnum\contacyrill=1     \def\cyrill{\ninewncyr}
  \def\cyrilb{\ninewncyb}  \def\cyrili{\ninewncyi}         \fi
  \textfont3=\nineex \scriptfont3=\sevenex \scriptscriptfont3=\sevenex
  \def\cmmib{\fam\cmmibfam\ninecmmib}  \textfont\cmmibfam=\ninecmmib
  \scriptfont\cmmibfam=\sixcmmib \scriptscriptfont\cmmibfam=\fivecmmib
  \def\cmbsy{\fam\cmbsyfam\ninecmbsy}  \textfont\cmbsyfam=\ninecmbsy
  \scriptfont\cmbsyfam=\sixcmbsy \scriptscriptfont\cmbsyfam=\fivecmbsy
  \def\cmcsc{\fam\cmcscfam\ninecmcsc} \scriptfont\cmcscfam=\eightcmcsc
  \textfont\cmcscfam=\ninecmcsc \scriptscriptfont\cmcscfam=\eightcmcsc
     \fi            \tt \ttglue=.5em plus.25em minus.15em
  \normalbaselineskip=11pt
  \setbox\strutbox=\hbox{\vrule height8pt depth3pt width0pt}
  \let\sc=\sevenrm \let\big=\ninebig \normalbaselines\rm}
\gdef\eightpoint{\def\rm{\fam0\eightrm}
  \textfont0=\eightrm \scriptfont0=\sixrm \scriptscriptfont0=\fiverm
  \textfont1=\eighti \scriptfont1=\sixi \scriptscriptfont1=\fivei
  \textfont2=\eightsy \scriptfont2=\sixsy \scriptscriptfont2=\fivesy
  \textfont3=\tenex \scriptfont3=\tenex \scriptscriptfont3=\tenex
  \def\mcal{\fam2 \eightsy}  \def\mmit{\fam1 \eighti}
  \textfont\itfam=\eightit \def\it{\fam\itfam\eightit}
  \textfont\slfam=\eightsl \def\sl{\fam\slfam\eightsl}
  \textfont\ttfam=\eighttt \scriptfont\ttfam=\eighttt
  \scriptscriptfont\ttfam=\eighttt \def\tt{\fam\ttfam\eighttt}
  \textfont\bffam=\eightbf \scriptfont\bffam=\sixbf
  \scriptscriptfont\bffam=\fivebf \def\bf{\fam\bffam\eightbf}
     \ifx\arisposta\amsrisposta   \ifnum\contaeuler=1
  \textfont\eufmfam=\eighteufm \scriptfont\eufmfam=\sixeufm
  \scriptscriptfont\eufmfam=\fiveeufm \def\eufm{\fam\eufmfam\eighteufm}
  \textfont\eufbfam=\eighteufb \scriptfont\eufbfam=\sixeufb
  \scriptscriptfont\eufbfam=\fiveeufb \def\eufb{\fam\eufbfam\eighteufb}
  \def\eurm{\eighteurm} \def\eurb{\eighteurb} \def\eusm{\eighteusm}
  \def\eusb{\eighteusb}       \fi    \ifnum\contaams=1
  \textfont\msamfam=\eightmsam \scriptfont\msamfam=\sixmsam
  \scriptscriptfont\msamfam=\fivemsam \def\msam{\fam\msamfam\eightmsam}
  \textfont\msbmfam=\eightmsbm \scriptfont\msbmfam=\sixmsbm
  \scriptscriptfont\msbmfam=\fivemsbm \def\msbm{\fam\msbmfam\eightmsbm}
     \fi       \ifnum\contacyrill=1     \def\cyrill{\eightwncyr}
  \def\cyrilb{\eightwncyb}  \def\cyrili{\eightwncyi}         \fi
  \textfont3=\eightex \scriptfont3=\sevenex \scriptscriptfont3=\sevenex
  \def\cmmib{\fam\cmmibfam\eightcmmib}  \textfont\cmmibfam=\eightcmmib
  \scriptfont\cmmibfam=\sixcmmib \scriptscriptfont\cmmibfam=\fivecmmib
  \def\cmbsy{\fam\cmbsyfam\eightcmbsy}  \textfont\cmbsyfam=\eightcmbsy
  \scriptfont\cmbsyfam=\sixcmbsy \scriptscriptfont\cmbsyfam=\fivecmbsy
  \def\cmcsc{\fam\cmcscfam\eightcmcsc} \scriptfont\cmcscfam=\eightcmcsc
  \textfont\cmcscfam=\eightcmcsc \scriptscriptfont\cmcscfam=\eightcmcsc
     \fi             \tt \ttglue=.5em plus.25em minus.15em
  \normalbaselineskip=9pt
  \setbox\strutbox=\hbox{\vrule height7pt depth2pt width0pt}
  \let\sc=\sixrm \let\big=\eightbig \normalbaselines\rm }
\gdef\tenbig#1{{\hbox{$\left#1\vbox to8.5pt{}\right.\n@space$}}}
\gdef\ninebig#1{{\hbox{$\textfont0=\tenrm\textfont2=\tensy
   \left#1\vbox to7.25pt{}\right.\n@space$}}}
\gdef\eightbig#1{{\hbox{$\textfont0=\ninerm\textfont2=\ninesy
   \left#1\vbox to6.5pt{}\right.\n@space$}}}
\def\alternativefont#1#2{\ifx\arisposta\amsrisposta \relax \else
\xdef#1{#2} \fi}
\global\contaeuler=0 \global\contacyrill=0 \global\contaams=0
%
%
%
%
\newbox\fotlinebb \newbox\hedlinebb \newbox\leftcolumn
\gdef\makeheadline{\vbox to 0pt{\vskip-22.5pt
     \fullline{\vbox to8.5pt{}\the\headline}\vss}\nointerlineskip}
\gdef\makehedlinebb{\vbox to 0pt{\vskip-22.5pt
     \fullline{\vbox to8.5pt{}\copy\hedlinebb\hfil
     \line{\hfill\the\headline\hfill}}\vss} \nointerlineskip}
\gdef\makefootline{\baselineskip=24pt \fullline{\the\footline}}
\gdef\makefotlinebb{\baselineskip=24pt
    \fullline{\copy\fotlinebb\hfil\line{\hfill\the\footline\hfill}}}
\gdef\doubleformat{\shipout\vbox{\Landspec\makehedlinebb
     \fullline{\box\leftcolumn\hfil\columnbox}\makefotlinebb}
     \advancepageno}
\gdef\columnbox{\leftline{\pagebody}}
\gdef\line#1{\hbox to\hsize{\hskip\leftskip#1\hskip\rightskip}}
\gdef\fullline#1{\hbox to\fullhsize{\hskip\leftskip{#1}%
\hskip\rightskip}}
\gdef\footnote#1{\let\@sf=\empty
         \ifhmode\edef\#sf{\spacefactor=\the\spacefactor}\/\fi
         #1\@sf\vfootnote{#1}}
\gdef\vfootnote#1{\insert\footins\bgroup
         \ifnum\dimnota=1  \eightpoint\fi
         \ifnum\dimnota=2  \ninepoint\fi
         \ifnum\dimnota=0  \tenpoint\fi
         \interlinepenalty=\interfootnotelinepenalty
         \splittopskip=\ht\strutbox
         \splitmaxdepth=\dp\strutbox \floatingpenalty=20000
         \leftskip=\oldssposta \rightskip=\olddsposta
         \spaceskip=0pt \xspaceskip=0pt
         \ifnum\sinnota=0   \textindent{#1}\fi
         \ifnum\sinnota=1   \item{#1}\fi
         \footstrut\futurelet\next\fo@t}
\gdef\fo@t{\ifcat\bgroup\noexpand\next \let\next\f@@t
             \else\let\next\f@t\fi \next}
\gdef\f@@t{\bgroup\aftergroup\@foot\let\next}
\gdef\f@t#1{#1\@foot} \gdef\@foot{\strut\egroup}
\gdef\footstrut{\vbox to\splittopskip{}}
\skip\footins=\bigskipamount
\count\footins=1000  \dimen\footins=8in
\catcode`@=12
\tenpoint
\ifnum\unoduecol=1 \hsize=\tothsize   \fullhsize=\tothsize \fi
\ifnum\unoduecol=2 \hsize=\collhsize  \fullhsize=\tothsize \fi
\global\let\lrcol=L      \ifnum\unoduecol=1
\output{\plainoutput{\ifnum\tipbnota=2 \clearnmbnota\fi}} \fi
\ifnum\unoduecol=2 \output{\if L\lrcol
     \global\setbox\leftcolumn=\columnbox
     \global\setbox\fotlinebb=\line{\hfill\the\footline\hfill}
     \global\setbox\hedlinebb=\line{\hfill\the\headline\hfill}
     \advancepageno  \global\let\lrcol=R
     \else  \doubleformat \global\let\lrcol=L \fi
     \ifnum\outputpenalty>-20000 \else\dosupereject\fi
     \ifnum\tipbnota=2\clearnmbnota\fi }\fi
\def\ifdoublepage{\ifnum\unoduecol=2 }
\gdef\yespagenumbers{\footline={\hss\tenrm\folio\hss}}
\gdef\ciao{ \ifnum\fdefcontre=1 \endfdef\fi
     \par\vfill\supereject \ifnum\unoduecol=2
     \if R\lrcol  \headline={}\nopagenumbers\null\vfill\eject
     \fi\fi \end}

\newskip\olddsposta \newskip\oldssposta
\global\oldssposta=\leftskip \global\olddsposta=\rightskip

\def\filldots{\leaders\hbox to 1em{\hss.\hss}\hfill}
\def\inquadrb#1 {\vbox {\hrule  \hbox{\vrule \vbox {\vskip .2cm
    \hbox {\ #1\ } \vskip .2cm } \vrule  }  \hrule} }
 \def\newline{\hfil\break}
\def\jump{\vskip\baselineskip} \newskip\iinnffrr
\def\sjump{\iinnffrr=\baselineskip
          \divide\iinnffrr by 2 \vskip\iinnffrr}
\def\bjump{\vskip\baselineskip \vskip\baselineskip}
\newcount\nmbnota  \def\clearnmbnota{\global\nmbnota=0}
\newcount\tipbnota \def\letterfootnote{\global\tipbnota=1}

\def\note#1{\global\advance\nmbnota by 1 \ifnum\tipbnota=1
    \footnote{$^{\rm\nttlett}$}{#1} \else {\ifnum\tipbnota=2
    \footnote{$^{\nttsymb}$}{#1}
    \else\footnote{$^{\the\nmbnota}$}{#1}\fi}\fi}
\def\nttlett{\ifcase\nmbnota \or a\or b\or c\or d\or e\or f\or
g\or h\or i\or j\or k\or l\or m\or n\or o\or p\or q\or r\or
s\or t\or u\or v\or w\or y\or x\or z\fi}
\def\nttsymb{\ifcase\nmbnota \or\dag\or\sharp\or\ddag\or\star\or
\natural\or\flat\or\clubsuit\or\diamondsuit\or\heartsuit
\or\spadesuit\fi}   \clearnmbnota
\def\numberfootnote{\global\tipbnota=0} \numberfootnote
\def\setnote#1{\expandafter\xdef\csname#1\endcsname{
\ifnum\tipbnota=1 {\rm\nttlett} \else {\ifnum\tipbnota=2
{\nttsymb} \else \the\nmbnota\fi}\fi} }
\newcount\nbmfig  \def\clearnbmfig{\global\nbmfig=0}
\gdef\figure{\global\advance\nbmfig by 1
      {\rm fig. \the\nbmfig}}   \clearnbmfig
\def\setfig#1{\expandafter\xdef\csname#1\endcsname{fig. \the\nbmfig}}
 \def\endformula{\eqno\numero $$}
 \def\efr{\endformula}
\newcount\frmcount \def\clearfrmcount{\global\frmcount=0}
\def\numero{\global\advance\frmcount by 1   \ifnum\indappcount=0
  {\ifnum\cpcount <1 {\hbox{\rm (\the\frmcount )}}  \else
  {\hbox{\rm (\the\cpcount .\the\frmcount )}} \fi}  \else
  {\hbox{\rm (\applett .\the\frmcount )}} \fi}
\def\nameformula#1{\global\advance\frmcount by 1%
\ifnum\draftnum=0  {\ifnum\indappcount=0%
{\ifnum\cpcount<1\xdef\spzzttrra{(\the\frmcount )}%
\else\xdef\spzzttrra{(\the\cpcount .\the\frmcount )}\fi}%
\else\xdef\spzzttrra{(\applett .\the\frmcount )}\fi}%
\else\xdef\spzzttrra{(#1)}\fi%
\expandafter\xdef\csname#1\endcsname{\spzzttrra}
\eqno \hbox{\rm\spzzttrra} $$}
\def\nfr{\nameformula}    
\def\nameali#1{\global\advance\frmcount by 1%
\ifnum\draftnum=0  {\ifnum\indappcount=0%
{\ifnum\cpcount<1\xdef\spzzttrra{(\the\frmcount )}%
\else\xdef\spzzttrra{(\the\cpcount .\the\frmcount )}\fi}%
\else\xdef\spzzttrra{(\applett .\the\frmcount )}\fi}%
\else\xdef\spzzttrra{(#1)}\fi%
\expandafter\xdef\csname#1\endcsname{\spzzttrra}
  \hbox{\rm\spzzttrra} }      \clearfrmcount
\newcount\cpcount \def\clearcpcount{\global\cpcount=0}
\newcount\subcpcount \def\clearsubcpcount{\global\subcpcount=0}
\newcount\appcount \def\clearappcount{\global\appcount=0}
\newcount\indappcount \def\clearindappcount{\indappcount=0}
\newcount\sottoparcount 

\def\applett{\ifcase\appcount  \or {A}\or {B}\or {C}\or
{D}\or {E}\or {F}\or {G}\or {H}\or {I}\or {J}\or {K}\or {L}\or
{M}\or {N}\or {O}\or {P}\or {Q}\or {R}\or {S}\or {T}\or {U}\or
{V}\or {W}\or {X}\or {Y}\or {Z}\fi    \ifnum\appcount<0
\immediate\write16 {Panda ERROR - Appendix: counter "appcount"
out of range}\fi  \ifnum\appcount>26  \immediate\write16 {Panda
ERROR - Appendix: counter "appcount" out of range}\fi}
\clearappcount  \clearindappcount \newcount\connttrre
\def\clearconnttrre{\global\connttrre=0} \newcount\countref
\def\clearcountref{\global\countref=0} \clearcountref
\def\chapter#1{\global\advance\cpcount by 1 \clearfrmcount
                 \goodbreak\null\vbox{\jump\nobreak
                 \clearsubcpcount\clearindappcount
                 \itemitem{\ttaarr\the\cpcount .\qquad}{\ttaarr #1}
                 \par\nobreak\jump\sjump}\nobreak}
\def\section#1{\global\advance\subcpcount by 1 \goodbreak\null
               \vbox{\sjump\nobreak\ifnum\indappcount=0
                 {\ifnum\cpcount=0 {\itemitem{\ppaarr
               .\the\subcpcount\quad\enskip\ }{\ppaarr #1}\par} \else
                 {\itemitem{\ppaarr\the\cpcount .\the\subcpcount\quad
                  \enskip\ }{\ppaarr #1} \par}  \fi}
                \else{\itemitem{\ppaarr\applett .\the\subcpcount\quad
                 \enskip\ }{\ppaarr #1}\par}\fi\nobreak\jump}\nobreak}
\clearsubcpcount
\def\appendix#1{\global\advance\appcount by 1 \clearfrmcount
                  \goodbreak\null\vbox{\jump\nobreak
                  \global\advance\indappcount by 1 \clearsubcpcount
          \itemitem{ }{\hskip-40pt\ttaarr Appendix\ #1}
             \nobreak\jump\sjump}\nobreak}
\clearappcount \clearindappcount
\def\references{\goodbreak\null\vbox{\jump\nobreak
   \itemitem{}{\ttaarr References} \nobreak\jump\sjump}\nobreak}

\clearcpcount\clearcountref

\def\setchap#1{\ifnum\indappcount=0{\ifnum\subcpcount=0%
\xdef\spzzttrra{\the\cpcount}%
\else\xdef\spzzttrra{\the\cpcount .\the\subcpcount}\fi}
\else{\ifnum\subcpcount=0 \xdef\spzzttrra{\applett}%
\else\xdef\spzzttrra{\applett .\the\subcpcount}\fi}\fi
\expandafter\xdef\csname#1\endcsname{\spzzttrra}}
\newcount\draftnum \newcount\ppora   \newcount\ppminuti
\global\ppora=\time   \global\ppminuti=\time
\global\divide\ppora by 60  \draftnum=\ppora
\multiply\draftnum by 60    \global\advance\ppminuti by -\draftnum
\def\droggi{\number\day /\number\month /\number\year\ \the\ppora
:\the\ppminuti}     \global\draftnum=0
\def\draftcomment#1{\ifnum\draftnum=0 \relax \else
{\ {\bf ***}\ #1\ {\bf ***}\ }\fi} 
%
%
\catcode`@=11
\gdef\Ref#1{\expandafter\ifx\csname @rrxx@#1\endcsname\relax%
{\global\advance\countref by 1    \ifnum\countref>200
\immediate\write16 {Panda ERROR - Ref: maximum number of references
exceeded}  \expandafter\xdef\csname @rrxx@#1\endcsname{0}\else
\expandafter\xdef\csname @rrxx@#1\endcsname{\the\countref}\fi}\fi
\ifnum\draftnum=0 \csname @rrxx@#1\endcsname \else#1\fi}
\gdef\beginref{\ifnum\draftnum=0  \gdef\Rref{\fairef}
\gdef\endref{\scriviref} \else\relax\fi
\ifx\risposta\mplarisposta \ninepoint \fi
\parskip 2pt plus.2pt \baselineskip=12pt}
\def\Reflab#1{[#1]} \gdef\Rref#1#2{\item{\Reflab{#1}}{#2}}
\gdef\endref{\relax}  \newcount\conttemp
\gdef\fairef#1#2{\expandafter\ifx\csname @rrxx@#1\endcsname\relax
{\global\conttemp=0 \immediate\write16 {Panda ERROR - Ref: reference
[#1] undefined}} \else
{\global\conttemp=\csname @rrxx@#1\endcsname } \fi
\global\advance\conttemp by 50  \global\setbox\conttemp=\hbox{#2} }
\gdef\scriviref{\clearconnttrre\conttemp=50
\loop\ifnum\connttrre<\countref \advance\conttemp by 1
\advance\connttrre by 1
\item{\Reflab{\the\connttrre}}{\unhcopy\conttemp} \repeat}
\clearcountref \clearconnttrre
\catcode`@=12
\ifx\risposta\mplarisposta \def\Reflab#1{#1.} \letterfootnote \fi

\def\slashchar#1{\setbox0=\hbox{$#1$} \dimen0=\wd0
     \setbox1=\hbox{/} \dimen1=\wd1 \ifdim\dimen0>\dimen1
      \rlap{\hbox to \dimen0{\hfil/\hfil}} #1 \else
      \rlap{\hbox to \dimen1{\hfil$#1$\hfil}} / \fi}
\ifx\oldchi\undefined \let\oldchi=\chi
  \def\cchi{{\raise 1pt\hbox{$\oldchi$}}} \let\chi=\cchi \fi

\def\frac#1#2{{\textstyle{#1 \over #2}}}

\def\half{\ifinner {\scriptstyle {1 \over 2}}\else {1 \over 2} \fi}

\def\simge{\rlap{\raise 2pt \hbox{$>$}}{\lower 2pt \hbox{$\sim$}}}
\def\simle{\rlap{\raise 2pt \hbox{$<$}}{\lower 2pt \hbox{$\sim$}}}

\def\vbig#1#2{{\vbigd@men=#2\divide\vbigd@men by 2%
\hbox{$\left#1\vbox to \vbigd@men{}\right.\n@space$}}}

%
%
\newcount\fdefcontre \newcount\fdefcount \newcount\indcount
\newread\filefdef  \newread\fileftmp  \newwrite\filefdef
\newwrite\fileftmp     \def\strip#1*.A {#1}
\def\futuredef#1{\beginfdef
\expandafter\ifx\csname#1\endcsname\relax%
{\immediate\write\fileftmp {#1*.A}
\immediate\write16 {Panda Warning - fdef: macro "#1" on page
\the\pageno \space undefined}
\ifnum\draftnum=0 \expandafter\xdef\csname#1\endcsname{(?)}
\else \expandafter\xdef\csname#1\endcsname{(#1)} \fi
\global\advance\fdefcount by 1}\fi   \csname#1\endcsname}

\def\beginfdef{\ifnum\fdefcontre=0
\immediate\openin\filefdef \jobname.fdef
\immediate\openout\fileftmp \jobname.ftmp
\global\fdefcontre=1  \ifeof\filefdef \immediate\write16 {Panda
WARNING - fdef: file \jobname.fdef not found, run TeX again}
\else \immediate\read\filefdef to\spzzttrra
\global\advance\fdefcount by \spzzttrra
\indcount=0      \loop\ifnum\indcount<\fdefcount
\advance\indcount by 1   \immediate\read\filefdef to\spezttrra
\immediate\read\filefdef to\sppzttrra
\edef\spzzttrra{\expandafter\strip\spezttrra}
\immediate\write\fileftmp {\spzzttrra *.A}
\expandafter\xdef\csname\spzzttrra\endcsname{\sppzttrra}
\repeat \fi \immediate\closein\filefdef \fi}
\def\endfdef{\immediate\closeout\fileftmp   \ifnum\fdefcount>0
\immediate\openin\fileftmp \jobname.ftmp
\immediate\openout\filefdef \jobname.fdef
\immediate\write\filefdef {\the\fdefcount}   \indcount=0
\loop\ifnum\indcount<\fdefcount    \advance\indcount by 1
\immediate\read\fileftmp to\spezttrra
\edef\spzzttrra{\expandafter\strip\spezttrra}
\immediate\write\filefdef{\spzzttrra *.A}
\edef\spezttrra{\string{\csname\spzzttrra\endcsname\string}}
\iwritel\filefdef{\spezttrra}
\repeat  \immediate\closein\fileftmp \immediate\closeout\filefdef
\immediate\write16 {Panda Warning - fdef: Label(s) may have changed,
re-run TeX to get them right}\fi}
\def\iwritel#1#2{\newlinechar=-1
{\newlinechar=`\ \immediate\write#1{#2}}\newlinechar=-1}
\global\fdefcontre=0 \global\fdefcount=0 \global\indcount=0
%
%
\null
%
%
%
%

\input psfig

%
\loadamsmath
\loadeuler
\mathchardef\bbalpha="710B
\mathchardef\bbbeta="710C
\mathchardef\bbgamma="710D
\mathchardef\bbxi="7118
\mathchardef\bbomega="7121
\mathchardef\sdir="2D6E
\mathchardef\dirs="2D6F
\def\balpha{{\bfmath\bbalpha}}

\def\bgamma{{\bfmath\bbgamma}}

\def\bxi{{\bfmath\bbxi}}
\def\U{{\rm U}}
\def\SU{{\rm SU}}
\pageno=0\baselineskip=14pt
\nopagenumbers{
\line{\hfill SWAT/96/112}
\line{\hfill CBPF-NF-024/96}
\line{\hfill\tt hep-th/9605069}
\line{\hfill Date}
\ifdoublepage \bjump\bjump\bjump\bjump\else\vfill\fi
\centerline{\capsone S-duality in $N=4$ supersymmetric gauge theories}
\centerline{\capsone with arbitrary gauge group}
\bjump\sjump
\centerline{\scaps Nicholas Dorey, Christophe Fraser, Timothy J. Hollowood}
\sjump
\centerline{\sl Department of Physics, University of Wales Swansea,}
\centerline{\sl Singleton Park, Swansea SA2 8PP, U.K.}
\centerline{\tt n.dorey, c.fraser, t.hollowood @swansea.ac.uk}
\sjump
\centerline{and}
\sjump
\centerline{\scaps Marco A. C. Kneipp}
\sjump
\centerline{\sl Centro Brasileiro de Pesquisas Fisicas (CBPF),
Rua Dr. Xavier Sigaud, 150}
\centerline{\sl 22290-180 Rio de Janeiro, Brazil}
\centerline{\tt kneipp@cbpfsu1.cat.cbpf.br}
\bjump\bjump
\ifdoublepage
\vfill
\eject\null\vfill\fi
\centerline{\capsone ABSTRACT}\sjump

The Goddard, Nuyts and Olive conjecture for electric-magnetic duality
in Yang-Mills theory with an arbitrary gauge group $G$ is extended by 
including a non-vanishing vacuum angle $\theta$. This extended
$S$-duality conjecture includes the case when the unbroken gauge
group is non-abelian and a definite prediction for the spectrum of
dyons results.

\sjump\vfill
\eject}
 \vfill

\yespagenumbers\pageno=1
%
%
In recent work 
[\Ref{SEN1},\Ref{POR1},\Ref{US},\Ref{SC},\Ref{GAU},\Ref{EW1},\Ref{GIB},\Ref{MM}], 
evidence has emerged that the electric-magnetic
duality conjectured by Goddard, Nuyts and Olive (GNO) [\Ref{GNO}] 
is an exact relation between $N=4$ supersymmetric gauge theories. 
In its original formulation, GNO duality has just one generator, which 
interchanges strong and weak coupling. 
This ordinary GNO duality requires that the spectrum of 
massive gauge bosons of a
gauge theory for a gauge group $G$ broken to $H$ by an adjoint Higgs
mechanism, is equal to the spectrum of magnetic monopoles for a dual
gauge theory based on a dual gauge group $G^*$ broken to $H^*$
[\Ref{GNO}]. In actual fact, the duality can only be exact in the
context of an $N=4$ supersymmetric gauge theory (unless additional
matter fields are added). The dual gauge
group is not necessarily isomorphic to the original group. For groups
with simply-laced Lie algebras $G^*\simeq G$; however for groups with 
non-simply-laced Lie algebras the groups $G\leftrightarrow G^*$
come in pairs: ${\rm SO}(2r+1)\leftrightarrow{\rm Sp}(r)$,
$F_4\leftrightarrow F_4'$ and $G_2\leftrightarrow G_2'$. In the latter
two cases the primes indicate that the corresponding dual groups are
related by an exchange of long and short roots.

In the minimal case, where $G=\SU(2)$, the inclusion of
non-zero vacuum (theta) angle leads to a larger group of duality
transformations on the parameters of the theory [\Ref{DUAL}];
namely the modular group ${\rm SL}(2,{\Bbb Z})$. 
In this letter, we will determine the corresponding duality group
which extends GNO duality to the case of non-zero theta angle for an arbitrary
gauge group. 
As in the case of gauge group $\SU(2)$, the additional symmetry comes
from the $\theta$ periodicity of the partition function. 
The spectrum of states must also be invariant under this shift, which
therefore provides a second symmetry generator.   
When combined with the original GNO duality, the new generator leads to an 
extended $S$-duality group which 
acts on an integer lattice of states. 
Our result is that this group in its most general form is a semi-direct product
of a subgroup of ${\rm SL}(2,{\Bbb Z})$ with the ordinary GNO
duality group ${\Bbb Z}_2$. The main goal of the paper will be to
elucidate the precise nature of this group. In particular, when the gauge
group is simply-laced, the $S$-duality group reduces to the modular
group itself; however, when the gauge group is non-simply-laced the
action of the $S$-duality group is more complicated, but there is nevertheless
a simple procedure for computing the conjectured spectrum of states.
There results an enhanced duality conjecture, where dyonic states of the theory
are conjectured to be gauge bosons of dual gauge theories with 
gauge group $G$ or $G^*$ in some characteristic pattern that 
we will elucidate. It remains a challenge to prove that these
states exist within the semi-classical approximation.
We should emphasize at this stage, however, that our picture
of extended duality does not require any additional conjectures over
and above the original GNO conjecture.

The exact duality that we are considering naturally occurs in the
context of an $N=4$ supersymmetric Yang-Mills theory with arbitrary
gauge group $G$.
We take all the fields to lie in a single sixteen dimensional supermultiplet.
All the fields transform in the adjoint representation of the
group and we take them to be Lie algebra valued. 
We can always work in a unitary gauge where the six
real scalar fields are constant on a large sphere at infinity.
The global ${\rm SO}(6)$ ${\cal R}$-symmetry is
spontaneously broken to  ${\rm SO}(5)$, and there are
consequently five massless Goldstone bosons. The spectrum of
massive states of the theory is completely
determined by considering only the remaining real Higgs field $\phi$,
arbitrarily chosen up to an ${\rm SO}(6)$ ${\cal R}$-symmetry
transformation. The bosonic part of the Lagrangian for 
this single scalar $\phi$ is then 
$$
{\cal L}
=-{e^2\over 32\pi}{\rm Im} \left[ \tau{\rm Tr}\left( F_{\mu\nu} + 
i { }^* F_{\mu\nu}
\right) \left( F^{\mu\nu} + i { }^* F^{\mu\nu}\right) \right] 
+ {1\over 2} {\rm Tr} \left( {\cal D}_\mu \phi
{\cal D}^\mu \phi \right) - V(\phi), 
\nfr{lagrangian}
where we use an othonormal basis for the algebra (${\rm Tr} \left( 
{\rm T}^{a} {\rm T}^{b} \right) = \delta^{ab}$) and  we have defined 
the complex coupling $\tau$:
$$
\tau = {\theta \over 2 \pi} + {4 \pi i \over e^2}.
\nfr{taudef}
The supersymmetric potential for $\phi$ vanishes 
which leads us naturally to the Prasad-Sommerfield limit $V(\phi) = 0$
[\Ref{BPS}].

Let $\phi_0$ be the constant Higgs field on the large
sphere at infinity in unitary gauge, chosen to lie within a 
Cartan subalgebra of the Lie algebra $g$: 
$$
\phi_0={\bfmath v}\cdot{\bfmath H},
\nfr{HF}
where $\bfmath H$ are the Cartan elements of $g$
considered as an $r={\rm rank}(g)$ vector. The simple roots 
${\balpha_i}$ of $g$ can always be chosen such that ${\bfmath v}\cdot
{\balpha_i}\geq 0 $. The Higgs field breaks the
symmetry to a subgroup $H\subset G$ 
which consists of group elements which commute with $\phi_0$:
$$
H=\left\{U\in G\vert\ U\phi_0U^{-1}=\phi_0\right\}.
\nfr{UGG}
Generically the unbroken gauge group will be the maximal torus of $G$;
however, if $\bfmath v$ is orthogonal to any simple root of $g$ then
the unbroken gauge group has a non-abelian component.
In general, therefore, $H$ is locally of the form $\U(1)^{r'}\times K$,
where $K$ is a semi-simple Lie group of rank $r-r'$. The global
definition of $H$, which requires the specification of a finite group,
will not be required in what follows. The Lie algebra $h$ of 
$H$ consists of the generators of $g$ commuting with
$\phi_0$. 

The evidence for GNO duality begins with the mass formulae of the
gauge bosons and monopoles in the theory.
Associating the gauge bosons with the Cartan-Weyl basis of the Lie
algebra $g$, 
the states corresponding to the Cartan elements are massless while the
states associated to the step generators $E_\balpha$ have a mass
$$
M_\balpha=e\vert{\bfmath v}\cdot\balpha \vert,
\nfr{MGB}
Massive gauge bosons are associated to the roots of $g$ with non-zero
inner product with ${\bfmath v}$. We will denote this subset of the
root system of $g$ as $\Phi'(g)$. The states
form multiplets of $K$ and carry
abelian charges with respect to the unbroken U$(1)^{r'}$.

Monopole solutions in these theories were found originally in [\Ref{BAIS}]
(see also [\Ref{EW2}]) by embeddings of the SU(2) monopole. As with
the gauge bosons, the monopole
solutions are associated to roots of the Lie algebra and their mass
spectrum is
$$
\tilde M_\balpha ={4\pi\over e}
\left\vert{\bfmath v}\cdot\balpha^*\right\vert,\ \ \
\balpha^*={\balpha\over\balpha^2}. 
\nfr{MMON}
where $\balpha\in\Phi'(g)$ for a non-trivial solution.
Notice that the spectrum of monopoles appears to be precisely equal to the
spectrum of massive gauge bosons in a dual theory with gauge coupling
$4\pi/\lambda e$ and gauge group $G^*$, whose Lie algebra $g^*$
has roots $\lambda\balpha/\balpha^2$,
where $\balpha$ are the roots of $g$, and $\lambda$ is a normalization 
constant. Actually this simplicity is somewhat illusory, as we discuss below. 
However, the above observation  
formed the original motivation for the GNO duality
conjecture. The normalization constant $\lambda$ is fixed to be
\note{In the following
$\vert\balpha_{\rm long}\vert$ and $\vert\balpha_{\rm short}\vert$ are
lengths of the long and short roots, respectively, of the Lie algebra
$g$. We will fix the normalization of the roots of the dual algebra
$g^*$ by demanding that its long and short roots have the same
length as those of $g$.}
$$
\lambda=\vert\balpha_{{\rm long}}\vert\vert\balpha_{{\rm short}}\vert.
\nfr{lamda} 
It will be convenient to define
$$
\eta={\vert\balpha_{\rm long}\vert^2\over\vert\balpha_{\rm short}\vert^2},
\efr
where
$$\eqalign{
\eta &= 1\ \ \ \hbox{   for the simply laced algebras }\  su(r), so(2r), 
e_6, e_7, e_8, \cr
\eta &= 2\ \ \ {\rm   for }\ so(2r+1), sp(r), f_4, \cr
\eta &= 3\ \ \ {\rm   for }\ g_2. \cr}
\efr

In the case of maximal symmetry breaking when $H={\rm U}(1)^r$
dramatic new evidence [\Ref{GAU},\Ref{EW1},\Ref{GIB}] 
for the GNO conjecture has been found, by
showing that a set of monopole states exists, within the semi-classical
approximation, with the mass spectrum given in \MMON. The non-trivial
observation is that monopoles associated to non-simple roots appear
as bound-states at threshold of monopoles associated to simple roots.
The situation with non-maximal symmetry breaking is not so clear.
The point is that the symmetry between \MGB\ and \MMON\ hides an
important difference. Remember that the 
gauge bosons form representations of $K$ which lead to 
degeneracies in the mass spectrum \MGB . Superficially it appears as
though these degeneracies are precisely mirrored in the 
monopole mass formula \MMON . However, it turns out that 
monopoles of equal mass are always part of a larger continuous
degeneracy of solutions. For instance, whenever two monopoles correspond 
to different roots $\balpha$ and $\bgamma$ which are related by a Weyl
group transformation of $k$, then they are both contained in the same 
connected manifold of solutions (the moduli space). 
In fact, even when two monopoles are degenerate but their associated roots are
not related by a Weyl group element, then there is a larger moduli
space which connects the two solutions. This subtlety occurs in
non-simply-laced cases for monopoles associated to short roots 
[\Ref{EW2},\Ref{EW3}]. This phenomena is referred to as an `accidental
degeneracy' since there is no apparant symmetry which relates the two
solutions. Hence the
degeneracy of states implied by \MMON\ 
is a pure illusion and in order to determine
the true degeneracy of monopole states one should presumably perform a
semi-classical quantization. In [\Ref{US}], some preliminary results
indicate that indeed the monopoles carry a degeneracy which is
consistent with GNO duality. For the present we shall simply assume
that GNO duality is correct and examine the consequences for the
spectrum of dyon states.

In the presence of non-zero $\theta$ angle, Witten [\Ref{WIT}] showed that
the Noether charge for the electric ${\rm U}(1)$ transformations generated
by the scalar field is
$$
N = {Q_e\over e}-{\theta e\over 8\pi^2}Q_m,
\nfr{NC}
where $Q_e$ and $Q_m$ are the total electric and magnetic charge 
of the classical field defined as the following surface integrals of
the electric and magnetic charges on the sphere at infinity
$$
Q_e={1\over\vert{\bfmath v}\vert}\int_{S^2_\infty}dS_i{\rm Tr}
\left(E_i\phi\right),\qquad
Q_m={1\over\vert{\bfmath v}\vert}\int_{S^2_\infty}dS_i{\rm Tr}
\left(B_i\phi\right).
\efr
The result \NC\ was derived in the context of
$\SU (2)$ gauge theory, but is in fact independent of the gauge group.
The GNO quantization condition states that the magnetic charge vector $\bxi_m$ 
has to be in the co-root lattice of $g$ 
which is  spanned by the duals of the simple roots $\balpha_{i}^*$, so
$$
Q_m = {4\pi\over e} \hat{\bfmath v}\cdot\bxi_m,
\efr
where $\hat{\bfmath v}={\bfmath v}/\vert{\bfmath v}\vert$.
The electric charge vector $\bxi_e$ lies in the
weight lattice of the representations under which the fields transform.
Since in this case all the fields are in the adjoint representation of $g$,
$\bxi_e$ has to lie in the root lattice of $g$
which is spanned by the simple roots $\balpha_i$, hence
$$
N = \hat{\bfmath v}\cdot\bxi_e.
\efr
This result is modified in the presence of matter
transforming under different representations of G, as is
the case in the finite $N=2$ Yang-Mills theories coupled
to fundamental hypermultiplets.
In the case at hand the Witten effect is recovered since, 
as in the $\SU (2)$ case, the monopoles acquire 
electric charge. The masses of Bogomol'nyi saturated states with a
given electric and magnetic charge is then $\vert{\bfmath v}
\vert\vert Q_e+iQ_m\vert$, as found by Osborn [\Ref{OS}].
Consequently, the mass formula for the monopoles is modified to
$$
\tilde M_\balpha =
e\left\vert\tau\left({\bfmath v}\cdot\balpha^*\right)\right\vert.
\efr
The universal mass formula for all Bogomol'nyi saturated states in a
theory with gauge group $G$ can now be written as
$$
M_G(X,\tilde\tau)
=\sqrt{4\pi\over{\rm Im}\tau}\left\vert{\bfmath v}\cdot(\bxi_e+\tau
\bxi_m)\right\vert,
\nfr{UMF}
where we have defined for later convenience
$$
\tilde\tau= {1\over\vert\balpha_{{\rm long}}\vert^2}
\left({\theta\over2\pi}+{4\pi i\over e^2}\right),\qquad\qquad
X=\pmatrix{\vert\balpha_{{\rm long}}\vert^{-1}\bxi_e\cr 
\vert\balpha_{{\rm long}}\vert\bxi_m\cr}.
\efr

The original GNO duality conjecture was made for the case of 
$\theta = 0$. It states that the monopoles of a theory with gauge
group $G$ can be thought of as the gauge bosons of an equivalent formulation of
the theory with a gauge group $G^*$ broken to $\U(1)^{r'}\times
K^*$ with a dual coupling constant  
$4\pi/\lambda e$. 
The conjecture was made on the basis of the mass formulae \MGB\ and
\MMON. For the present purposes, 
we will assume GNO duality is an exact relation between
theories and deduce the larger duality group which ensues when
a non-zero vacuum angle is included. 

We now consider the $\theta$ dependence of the action. 
The term $F^*F$ can be written as
a total derivative, so that $\int {\rm Tr} (F^*F)$ is a function
of the gauge field on the large sphere at infinity $S^3_{\infty}$.
In fact, it is proportional to the winding number of the gauge field $A_\mu$
where $A_\mu$ maps $S^3_{\infty}$ into the lie algebra $g$ of $G$.
It is known that for general gauge group\note{MACK would like to thank
J. Labastida for providing a proof of the following result.
The proof may also be obtained from the formulae in [\Ref{GO}].}
$$
{e^2\over 32\pi^2}\int {\rm Tr} (F^*F) = {N\over\vert\balpha_{\rm long}\vert^2}
\efr
where N is the integer winding number of the gauge field $A_\mu$.

The partition function involves a sum over all integers $N$,
implying that the $\theta$
periodicity of the partition function is simply
$$
\theta \rightarrow \theta + 2 \pi\vert \balpha_{\rm long}\vert^2.
\nfr{thetaperiod}
In terms of the complex coupling $\tilde\tau$ we have
$$
\tilde\tau \rightarrow \tilde\tau + 1.
\nfr{tauperiod}

In order to define the extended duality group of the theory
we combine the $\theta$-periodicity of the partition function,
which we refer to as ${\bfmath T}$, with the conjectured
GNO duality, which we refer to as ${\bfmath S}$.
Consider the action of the two generators $\bfmath S$ and $\bfmath T$
on these new electric and magnetic charge vectors, and on the complex
coupling $\tilde\tau$:
$$
\eqalign{
{\bfmath S}:\qquad
&\tilde\tau \mapsto - {1\over\eta\tilde\tau}, \cr
{\bfmath T}:\qquad
&\tilde\tau \mapsto \tilde\tau + 1, \cr} \qquad
\eqalign{
X&\mapsto
\pmatrix{ 0&1/\sqrt\eta\cr -\sqrt\eta&0\cr}X,\cr 
X&\mapsto
\pmatrix{1&-1\cr 0&1\cr}X. \cr}
\nfr{dualS}
Furthermore these transformation have an action on the gauge group of
the theory
$$\eqalign{
{\bfmath S}:\qquad &G \mapsto G^*, \cr
{\bfmath T}:\qquad &G \mapsto G. \cr}
\efr
It can easily be verified that these transformations are a symmetry of
the universal mass formula \UMF:
$$
M_{{\bfmath S}G}({\bfmath S}X,{\bfmath
S}\tilde\tau)=M_G(X,\tilde\tau),\qquad
M_{{\bfmath T}G}({\bfmath T}X,{\bfmath T}\tilde\tau)=M_G(X,\tilde\tau).
\nfr{MSYM}

If $G$ is simply-laced ($\eta=1$ and $G \simeq G^*$) 
then ${\bfmath S}$ and ${\bfmath T}$ generate the modular group
${\rm SL}(2,{\Bbb Z})$, and we recover the standard extended
duality conjecture. On the contrary, if 
G is non-simply-laced ($\eta \ne 1$) then ${\bfmath S}$ is not
a modular transformation and it is helpful to
consider some additional generators, which although redundant, help
elucidate the action of the duality group.
This is done by separating out the group of transformations which
relate multiplets transforming under the same gauge group.
This subgroup of the full duality group will be generated by
${\bfmath T}$, ${\bfmath STS}$ and ${\bfmath S}^2$, where 
$$
{\bfmath STS}:\qquad G \mapsto G, \qquad
\tilde\tau \mapsto {\tilde\tau\over 1 - \eta\tilde\tau},\qquad
X \mapsto \pmatrix{-1&0 \cr -\eta&-1\cr}X. 
\efr
Notice that although the transformation
${\bfmath S}^2$ acts trivially on the complex coupling, it
reverses the sign of the electric and magnetic charges (it
is the $CP$ operator for the theory). 
${\bfmath T}$, ${\bfmath STS}$, and ${\bfmath S}^2$
generate a subgroup of the modular group ${\rm SL}(2,{\Bbb Z})$
called $\Gamma_0\left(\eta\right)$.\note{We use the terminology for
subgroups of the modular group defined in [\Ref{Schoeneberg}]. For a discussion
of the subgroups of ${\rm SL}(2,{\Bbb Z})$ in $N=2$ theory see for example 
[\Ref{Minahan}]. The appearance of $\Gamma_0(2)$ for the case ${\rm
Sp(n)}\leftrightarrow{\rm SO}(2n+1)$ was first noticed in
[\Ref{GAU}]. For a related discussion of the duality groups of
theories with arbitrary gauge groups see [\Ref{POR2}].}

A general transformation in $\Gamma_0 (\eta)$ has the form
$$
\tilde\tau \mapsto {a\tilde\tau+b\over c\tilde\tau+d}, \qquad
X\mapsto \pmatrix{a&-b\cr -c&d\cr}X,
\efr
where $a,b,c,d$ are integers such that $ad-bc=1$ and $c=0$ modulo $\eta$.

Now we identify the extended group of duality transformations. 
The transformation
${\bfmath S}$ generates ${\Bbb Z}_4$, however the ${\Bbb Z}_2$
subgroup generated by ${\bfmath S}^2$ is already a subgroup of
$\Gamma_0(\eta)$. Hence, the full $S$-duality group $\cal D$ of the theory is
the ${\Bbb Z}_2$ quotient of a semi-direct product:
$$
{\cal D} = \left[\Gamma_0 (\eta ) {\Bbb \dirs} {\Bbb Z}_4\right]/{\Bbb
Z}_2,
\efr
generated by ${\bfmath S}$ and ${\bfmath T}$. Notice that 
this is isomorphic to ${\rm SL}(2,{\Bbb Z})$ when $\eta=1$. 
The universal mass formula \UMF\ is now invariant under any transformation in
${\bfmath U}\in\cal D$:
$$
M_{{\bfmath U}G}({\bfmath U}X,{\bfmath U}\tilde\tau)=M_G(X,\tilde\tau).
\efr 

Given the duality symmetry established above we can determine the
spectrum of states. First of all, the spectrum of massive gauge
bosons is
$$
M_G\left(X_\balpha,\tilde\tau\right),\qquad\balpha\in\Phi'(g),
\efr
where
$$
X_\balpha=\pmatrix{\vert\balpha_{\rm
long}\vert^{-1}\balpha\cr 0\cr}.
\efr
The extended duality states that for each element ${\bfmath U}
\in{\cal D}$ there
is a reformulation of the theory with gauge group ${\bfmath U}G$ and with 
coupling constant ${\bfmath U}\tilde\tau$. The spectrum of the theory
must contain the gauge bosons of each of the dual formulations,
i.e. the spectrum of the theory must contain states of mass
$$
M_{{\bfmath U}G}(X_{\tilde\balpha},{\bfmath U}\tilde\tau),
\qquad\tilde\balpha\in\Phi'({\bfmath U}g),
\efr
for each ${\bfmath U}\in{\cal D}$.
By using the symmetry of the mass formula \MSYM\ the spectrum is
equivalently
$$
M_G({\bfmath U}^{-1}X_{\tilde\balpha},\tilde\tau),
\qquad\tilde\balpha\in\Phi'({\bfmath U}g),
\nfr{SPEC}
for each ${\bfmath U}\in{\cal D}$. To find the spectrum explicitly we
note the states can be split into two sets. The first set is generated
by ${\bfmath U}={\bfmath A}$, where ${\bfmath A}\in\Gamma_0(\eta)$,
i.e. have charge vectors 
$$
{\bfmath A}^{-1}X_\balpha=
\pmatrix{p\vert\balpha_{\rm
long}\vert^{-1}\balpha\cr 
q\vert\balpha_{\rm long}\vert^{-1}\balpha\cr},\qquad\qquad\balpha\in\Phi'(g),
\efr
where $q$ and $p$ are co-prime integers and $q=0$ modulo $\eta$, and hence
have masses
$$
\sqrt{4\pi\over{\rm Im}\tau}\left\vert\left(p+q\vert\balpha_{\rm
long}\vert^{-2}\tau\right)\bfmath v\cdot\balpha\right\vert,\qquad\qquad\balpha
\in\Phi'(g).
\nfr{MFS}
The second set of states is generated by ${\bfmath U}={\bfmath
S}{\bfmath A}$, where ${\bfmath A}\in\Gamma_0(\eta)$, 
i.e. have charge vectors 
$$
{\bfmath A}^{-1}{\bfmath S}^{-1}X_{\lambda\balpha^*}=
\pmatrix{p\vert\balpha_{\rm long}\vert\balpha^*\cr 
q\vert\balpha_{\rm
long}\vert\balpha^*\cr},\qquad\qquad\balpha\in\Phi'(g),
\efr
where $q$ and $p$ are co-prime integers and $q\neq0$ modulo $\eta$, and hence
have masses
$$
\sqrt{4\pi\over{\rm Im}\tau}\left\vert\left(p\vert\balpha_{\rm
long}\vert^2+q\tau\right)\bfmath v\cdot\balpha^*\right\vert,
\qquad\qquad\balpha\in\Phi'(g).
\nfr{MSS}
For example, the spectrum of monopoles \MMON\ is recovered by taking
$(p,q)=(0,1)$ in \MSS. 

So the complete mass spectrum can be described as follows. States are
associated to the co-prime pair of integers $(p,q)$ familiar from the
$\SU(2)$ theory. If $q=0$ modulo
$\eta$ then the states have masses given by \MFS\ and transform in
representations of $K$, isomorphic to those of the gauge bosons of the
$G$ theory. On the contrary if $q\neq0$ modulo $\eta$,
then the states have masses given by \MSS\ and transform in
representations of $K^*$, isomorphic to the gauge bosons of the
$G^*$ theory. We have illustrated the lattice of states that arise
for the case $\eta=2$ in figure 1, and for $\eta=3$ in figure 2. 
At each node $(p,q)$ labelled by
$G$, or $G^*$, there are a set of states with masses \MFS, or \MSS,
respectively. Clearly when $\eta=1$ 
the group is self-dual and picture is
directly analogous to the $SL(2,{\Bbb Z})$ lattice of states in the
$\SU(2)$ theory. We should emphasize that in the general case, unlike
the SU(2) example, the lengths of vectors on the lattices do not encode
the masses of the states; rather one must apply the mass formulae
\MFS\ and \MSS.

\vbox{\centerline{\psfig{figure=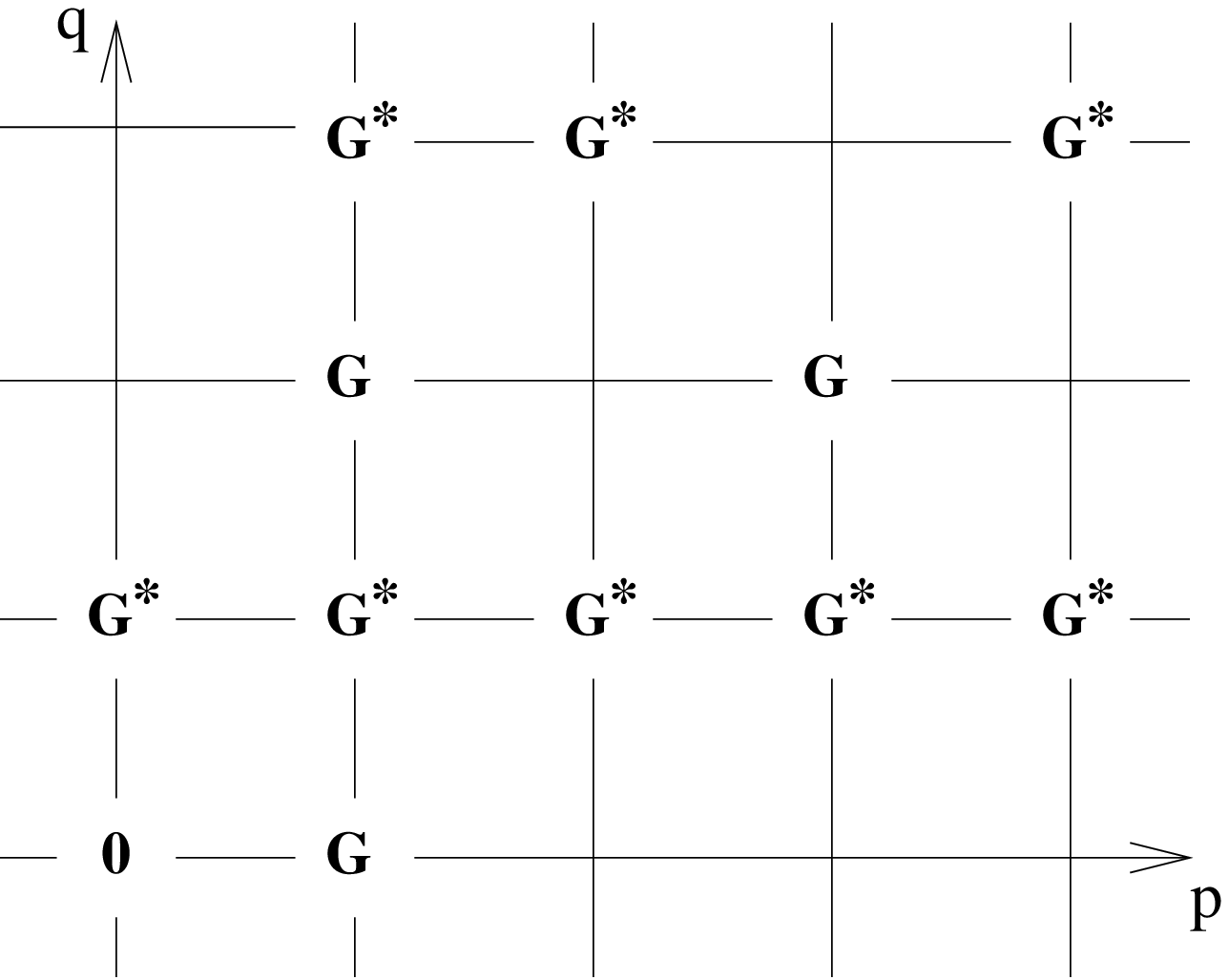,height=6cm}}
\bjump
\centerline{Figure 1. The lattice of states for $\eta=2$.}}

\bjump
\vbox{\centerline{\psfig{figure=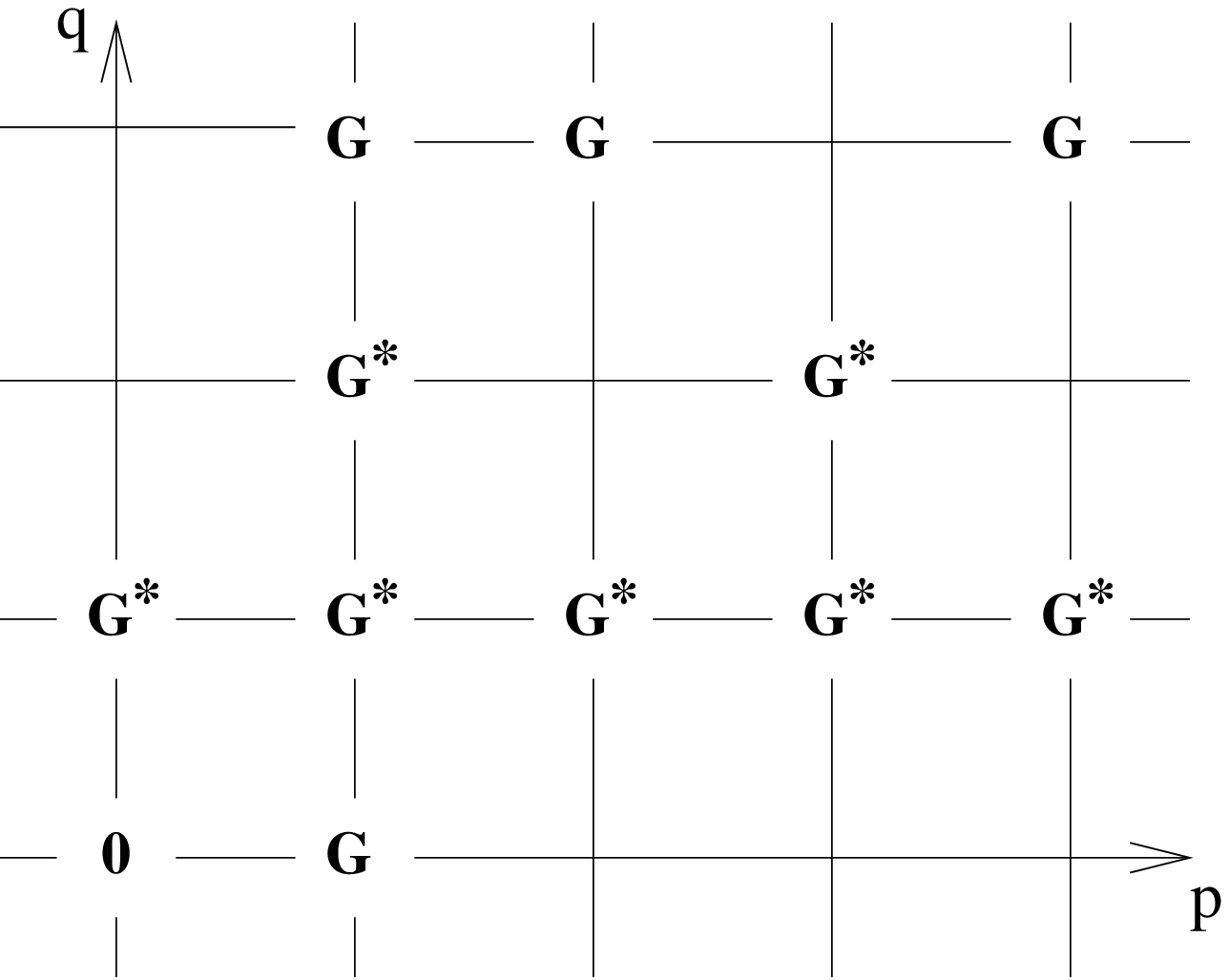,height=6cm}}
\bjump
\centerline{Figure 2. The lattice of states for $\eta=3$.}}

\bjump
We have seen that, starting from GNO duality, 
the required $\theta$-periodicity of the spectrum dictates a unique
S-duality group. Hence the GNO duality conjecture leads directly to
a corresponding S-duality conjecture. As in the SU(2) theory, a strong
test of this conjecture is the existence at the semi-classical level of
the predicted spectrum of dyons. 
In particular, the conjecture predicts a tower of dyon states with the
same set of magnetic charges as the monopoles themselves,
by taking $(p,q)=(p,\eta)$ (with $p$ and $\eta$ co-prime) in \MFS, for
short roots only, giving masses
$$
\sqrt{4\pi\over{\rm Im}\tau}
\left\vert\left(p\vert\balpha_{\rm short}\vert^2+\tau\right){\bfmath
v}\cdot\balpha^*\right\vert,\qquad\balpha\in\Phi_{\rm
short}'(g),\qquad (p,\eta)\ \hbox{co-prime},
\nfr{JZO}
and $(p,q)=(p,1)$ in \MSS, for any roots, giving masses
$$
\sqrt{4\pi\over{\rm Im}\tau}
\left\vert\left(p\vert\balpha_{\rm long}\vert^2+\tau\right){\bfmath
v}\cdot\balpha^*\right\vert,\qquad\balpha\in\Phi'(g).
\nfr{JZT}
The states in \JZO\ and \JZT\ are the
analogues of the Julia-Zee dyons in the $\SU(2)$ theory. These 
states should be obtained, in semi-classical limit, by quantizing the
$\U(1)$ degree-of-freedom associated with electric charge rotations of
a BPS monopole [\Ref{TW}]. 

CF would like to thank Martin Groves for useful discussions.
ND and TJH are supported by PPARC Advanced Fellowships. 
MACK is supported by a CNPq fellowship.

\references

\beginref
\Rref{GNO}{P. Goddard, J. Nuyts and D. Olive, Nucl. Phys. {\bf B125}
(1977) 1}
\Rref{US}{N. Dorey, C.Fraser, T.J. Hollowood and M.A.C. Kneipp, 
`Non-Abelian Duality in N=4 Supersymmetric Gauge theories' {\tt hep-th/9512116}}
\Rref{SEN1}{C. Vafa and E. Witten, Nucl. Phys. {\bf B431} (1994) 3-77; A. Sen, 
Phys. Lett. {\bf B329} (1994) 217-221}
\Rref{POR1}{M. Porrati, `On the Existence of States Saturating the 
Bogomol'nyi Bound in $N=4$ Supersymmetry', {\tt hep-th/9505187}}
\Rref{SC}{S.A. Connell, `The Dynamics of the ${\rm SU}(3)$ $(1,1)$
Magnetic Monopole', Unpublished preprint available by anonymous
ftp from:\newline {\tt <ftp://maths.adelaide.edu.au/pure/mmurray/oneone.tex>}}
\Rref{GAU}{J.P. Gauntlett and D. A. Lowe, `Dyons and S-duality in 
$N=4$ Supersymmetric Gauge Theory', {\tt hep-th/9601085}}
\Rref{GIB}{G.W. Gibbons, `The Sen conjecture for fundamental monopoles of
distinct types', {\tt hep-th/9603176}} 
\Rref{EW1}{K. Lee, E.J. Weinberg and P. Yi, `Electromagnetic Duality
and $\SU(3)$ Monopoles', {\tt hep-th/9601097}, `The Moduli Space
of Many BPS Monopoles for Arbitrary Gauge Groups', {\tt hep-th/9602167}}
\Rref{MM}{M.K. Murray, `A Note on the $(1,1,\dots ,1)$ Monopole Metric', {\tt hep-th/9605054}}
\Rref{BPS}{E.B. Bogomol'nyi, Sov. J. Nucl. Phys. {\bf 24} (1976);
 M.K. Prasad and C.M. Sommerfield, Phys. Rev. Lett. {\bf 35}
(1975) 760}
\Rref{DUAL}{J. Cardy and E. Rabinovici, Nucl. Phys. {\bf B205}
(1982) 1; J. Cardy, Nucl. Phys. {\bf B205} (1982) 17; A. Shapere
and F. Wilczek, Nucl. Phys. {\bf B320} (1989) 669; A. Font,
L. Ibanez, D. Lust and F. Quevedo, Phys. Lett.  {\bf B249}
(1990) 35.}
\Rref{BAIS}{F.A. Bais, Phys. Rev. {\bf D18} (1978) 1206}
\Rref{EW2}{E.J. Weinberg, Phys. Rev. {\bf D20} (1979) 936;
Nucl. Phys. {\bf B167} (1980) 500; Nucl. Phys. {\bf B203} (1982) 445}
\Rref{WIT}{E. Witten, Phys. Lett. {\bf B86} (1979) 283}
\Rref{Schoeneberg}{B. Schoeneberg, {\it Elliptic Modular Functions: An 
Introduction}, 
Springer-Verlag 1974}
\Rref{Minahan}{J.A. Minahan and D. Nemeschansky, `$N=2$ Super Yang-Mills and
Subgroups of ${\rm SL}(2,{\Bbb Z})$' {\tt hep-th/9601059}}
\Rref{POR2}{L. Girardello, A. Giveon, M. Porrati and A. Zaffaroni,
`S-Duality in $N=4$ Yang-Mills Theories' {\tt hep-th/9507064};
Nucl. Phys. {\bf B448} (1995) 127; Phys. Lett. {\bf B334} (1994) 331}
\Rref{TW}{E. Tomboulis and G. Woo, Nucl. Phys. {\bf B107} (1976) 221}
\Rref{EW3}{E.J. Weinberg, Phys. Lett. {\bf B119} (1982) 151\newline
A. Abdouelasood, Phys. Lett. {\bf B137} (1984) 77}
\Rref{GO}{P. Goddard and D. Olive, Int. J. Mod. Phys. {\bf A1} (1986)
303}
\Rref{OS}{H. Osborn, Phys. Lett. {\bf B83} (1979) 321}
\endref

\ciao